\documentclass[aps,twocolumn,showpacs,superscriptaddress,preprintnumbers,amsmath,amssymb,pre,longbibliography,floatfix]{revtex4-1}
\usepackage{graphicx}
\usepackage{amsmath}
\usepackage{amssymb}
\usepackage{mathtools}
\usepackage{makeidx}
\usepackage{amsfonts}
\usepackage{bm}
\usepackage{dcolumn}
\usepackage{color}
\usepackage{xcolor}
\usepackage{xfrac}
\usepackage{comment}
\usepackage{wasysym}
\usepackage{bbold}
\usepackage[colorlinks,
linkcolor=blue,
anchorcolor=blue,
citecolor=blue,
urlcolor=blue
]{hyperref}
\usepackage[normalem]{ulem}

\graphicspath{{./figures/}}

\begin{document}

\newcommand{\noteJS}[1]{\textcolor{red}{#1}}
\newcommand{\nwc}{\newcommand}
\nwc{\vs}{\vspace}
\nwc{\hs}{\hspace}
\nwc{\la}{\langle}
\nwc{\ra}{\rangle}
\nwc{\nn}{\nonumber}
\nwc{\Ra}{\Rightarrow}
\nwc{\wt}{\widetilde}
\nwc{\lw}{\linewidth}
\nwc{\ft}{\frametitle}
\nwc{\ben}{\begin{enumerate}}
\nwc{\een}{\end{enumerate}}
\nwc{\bit}{\begin{itemize}}
\nwc{\eit}{\end{itemize}}
\nwc{\dg}{\dagger}
\nwc{\mA}{\mathcal A}
\nwc{\mD}{\mathcal D}
\nwc{\mB}{\mathcal B}

\nwc{\Tr}[1]{\underset{#1}{\mbox{Tr}}~}
\nwc{\D}[2]{\frac{d #1}{d #2}}
\nwc{\pd}[2]{\frac{\partial #1}{\partial #2}}
\nwc{\ppd}[2]{\frac{\partial^2 #1}{\partial #2^2}}
\nwc{\fd}[2]{\frac{\delta #1}{\delta #2}}
\nwc{\pr}[2]{K(i_{#1},\alpha_{#1}|i_{#2},\alpha_{#2})}
\nwc{\av}[1]{\left< #1\right>}
\nwc{\alert}[1]{\textcolor{red}{#1}}
\nwc{\PSP}[1]{\textcolor{blue}{#1}}

\nwc{\zprl}[3]{Phys. Rev. Lett. ~{\bf #1},~#2~(#3)}
\nwc{\zpre}[3]{Phys. Rev. E ~{\bf #1},~#2~(#3)}
\nwc{\zpra}[3]{Phys. Rev. A ~{\bf #1},~#2~(#3)}
\nwc{\zjsm}[3]{J. Stat. Mech. ~{\bf #1},~#2~(#3)}
\nwc{\zepjb}[3]{Eur. Phys. J. B ~{\bf #1},~#2~(#3)}
\nwc{\zrmp}[3]{Rev. Mod. Phys. ~{\bf #1},~#2~(#3)}
\nwc{\zepl}[3]{Europhys. Lett. ~{\bf #1},~#2~(#3)}
\nwc{\zjsp}[3]{J. Stat. Phys. ~{\bf #1},~#2~(#3)}
\nwc{\zptps}[3]{Prog. Theor. Phys. Suppl. ~{\bf #1},~#2~(#3)}
\nwc{\zpt}[3]{Physics Today ~{\bf #1},~#2~(#3)}
\nwc{\zap}[3]{Adv. Phys. ~{\bf #1},~#2~(#3)}
\nwc{\zjpcm}[3]{J. Phys. Condens. Matter ~{\bf #1},~#2~(#3)}
\nwc{\zjpa}[3]{J. Phys. A ~{\bf #1},~#2~(#3)}
\nwc{\zpjp}[3]{Pramana J. Phys. ~{\bf #1},~#2~(#3)}

\title{Is active motion beneficial for target search with resetting in a thermal environment?}
\author{P. S. Pal}
\affiliation{School of Physics, Korea Institute for Advanced Study, Seoul 02455, Korea}
\author{Jong-Min Park}
\affiliation{Asia Pacific Center for Theoretical Physics, Pohang, 37673, Republic of Korea}
\affiliation{Department of Physics, Postech, Pohang 37673, Republic of Korea}
\author{Arnab Pal}
\affiliation{The Institute of Mathematical Sciences, CIT Campus, Taramani, Chennai 600113, India}
\affiliation{Homi Bhabha National Institute, Training School Complex, Anushakti Nagar, Mumbai 400094, India}
\author{Hyunggyu Park}
\affiliation{Quantum Universe Center, Korea Institute for Advanced Study, Seoul 02455, Korea}
\author{Jae Sung Lee}
\affiliation{School of Physics, Korea Institute for Advanced Study, Seoul 02455, Korea}
\begin{abstract}
Stochastic resetting has recently emerged as an efficient target-searching strategy in various physical and biological systems. The efficiency of this strategy depends on the type of environmental noise, whether it is thermal or telegraphic (active). While the impact of each noise type on a search process has been investigated separately, their combined effects have not been explored.
In this work, we explore the effects of stochastic resetting on an active system, namely a self-propelled run-and-tumble particle immersed in a thermal bath. In particular, we assume that the position of the particle is reset at a fixed rate with or without reversing the direction of self-propelled velocity. Using standard renewal techniques, we compute the mean search time of this active particle to a fixed target and investigate the interplay between active and thermal fluctuations. We find that the active search can outperform the Brownian search when the magnitude and flipping rate of self-propelled velocity are large and the strength of environmental noise is small. Notably, we find that the presence of thermal noise in the environment helps reduce the mean first passage time of the run-and-tumble particle compared to the absence of thermal noise. Finally, we observe that reversing the direction of self-propelled velocity while resetting can also reduce the overall search time. 
\end{abstract}
\maketitle
\section{Introduction}
\label{intro}

Target search is essential for the survival of biological organisms. Notable examples include searching for energy sources, food, shelter, and suitable mates.
Searching processes are also prevalent in cells. For instance, transcript factors search for target sequences along DNA strands through a process known as \textit{facilitated diffusion} ~\cite{Winter, Berg, Benichou}, which involves one-dimensional diffusion and stochastic jumps to other locations.
These searching processes are basically conducted by agents that stochastically wander their environment until they reach the target. Therefore, the searching time of the processes depends on their strategies how to wander around for searching. The representative measure for this searching time is the mean first passage time (FPT). Thus, one of the primary concerns in studies related to the searching problem is to find an optimal wandering strategy that minimizes the mean FPT.  

In the past decade, stochastic resetting has emerged as an efficient searching strategy that reduces the mean FPT~\cite{evans2011,gupta2014,evans2019,majumdar2015,eule2015,pal-ness,mendez2016,basu2019,olsen2024,reuveni2016,pal2017,pal2019prl,kusmirez2014,chechkin2018,belan2018}. Stochastic resetting is a renewal process where a dynamic system is interrupted at random times and reset to a predetermined configuration. Initially applied to a one-dimensional freely diffusing Brownian particle searching for a specific target, it has been demonstrated that stochastic resetting eliminates the possibility of the particle straying far from the target, resulting in a finite mean FPT. In contrast, the mean FPT for a free diffusing particle without resetting diverges. Subsequently, researchers have extensively studied various aspects of stochastic resetting, including properties of its nonequilibrium steady state~\cite{evans2011,pal-ness}, methods to further accelerate search processes~\cite{evans2011,reuveni2016,pal2017} and thermodynamic prospect of the search process~\cite{psp2023}. Beyond  its applications in physics, stochastic resetting also plays a crucial role in accelerating various biological processes. Examples include kinetic proofreading ~\cite{BarZiv,Murugan}, protein-folding process ~\cite{Bhaskaran, Hyeon, Chakrabarti}, and chemical reaction process ~\cite{unbinding}. See \cite{evans2019,pal22,pal23,pal23_sm} for an extensive review on the topic. 

However, the impact of stochastic resetting on biological organisms is unclear since their dynamics are often more complex than simple diffusion. In fact, many biological searching agents consume energy and exhibit self-propelled motion, known as active particles. These active particles operate in a nonequilibrium regime and therefore do not follow the fluctuation-dissipation relation. Examples of active particles include motile cells, motor proteins~\cite{Alberts-book}, and artificial active Brownian particles such as self-propelled colloids~\cite{Howse07,Zheng13}, nano-rotors~\cite{Nourhani13}, vibrated granular particles~\cite{Feitosa04,Kumar11}. Various models have been proposed to describe the motions of active particles, such as run-and-tumble particles (RTP)~\cite{Schnitzer93,Malakar18}, active Ornstein-Uhlenbeck particles (AOUP)~\cite{Bonilla19}, and active Brownian particles (ABP)~\cite{Romanczuk12}. These active motions give rise to unique thermodynamic properties, which have never been observed in passive systems~\cite{Krishnamurthy, Zakine, JSLee2020, JSLee2022}. 

For examining the impact of active particle on the search process, in this study, we focus on a target-searching process by RTP particles under stochastic resetting dynamics. Earlier researches in this area have primarily concentrated on the dynamics of RTP without considering the influence of the environment~\cite{Evans18}. This RTP-only dynamics can be modeled by substituting the thermal random noise with telegraphic noise~\cite{Rosenau93}. In reality, however, active particles usually exist in environments that provide thermal noise and influence their dynamics. Hence, it is important to study the combined effects of the active motion and thermal noise to gain a comprehensive understanding of search dynamics in biological systems.  

Here, we study the dynamics of a RTP that moves freely in one-dimensional space and is reset to a predetermined position at random times with a constant rate in the presence of thermal noise. Employing generic renewal techniques, we compute  the moment generating function of FPT, then evaluate mean FPT and finally, investigate the combined effect of thermal noise and active motion on this target-search process. The results clearly verify that active motion can reduce the mean FPT compared to that of Brownian motion within a certain range of parameters. More specifically, our comprehensive analysis reveals that an active particle can outperform a passive Brownian particle when the magnitude and flipping rate of the self-propelled speed are high and the strength of thermal noise is low. Additionally, we demonstrate that two other factors, the presence of a thermal environment and reversing the direction of the self-propelled velocity during resetting, are helpful in reducing the search time of the RTP.

The rest of this paper is organized as follows: In Section~\ref{sec:model}, we provide a detailed description of our system setup, including the governing dynamics of the RTP and resetting strategy used in the search process. Section~\ref{sec:mgf} introduces the moment generating function used to calculate the mean FPT. The results are presented in Section~\ref{sec:results}, which is divided into three subsections: Section~\ref{sec:unconditional_FTP} demonstrates the effect of active motion on mean FPT, Section~\ref{sec:effect_noise_eta} discusses the effects of thermal noise and reversal of active-motion direction during resetting on the mean FPT, and Section~\ref{sec:optimal} presents comprehensive analysis results on when an RTP outperforms a Brownian particle as a search agent. Finally, we summarize our findings in Section~\ref{sec:conclusion}.

\section{Model}
\label{sec:model}
We consider a RTP that diffuses freely in one-dimensional space. Run-and-tumble dynamics is modeled by an active speed $f$ that randomly reverses its direction at a flipping rate $r_{\rm F}$. The particle is situated in a thermal environment characterized by a diffusion constant $D$, which is connected to the environmental temperature $T$ by the relation $D=\gamma^{-1}k_{\rm B}T$, where $k_{\rm B}$ is the Boltzmann constant. Then, the motion of the particle is described by the following Langevin equation:
\begin{align} \label{eq:RTP_equation}
    \dot x&=\sigma(t)f+\xi(t),
\end{align}
where $\xi(t)$ is a Gaussian white noise with zero mean and $\la \xi(t)\xi(t')=2D\delta(t-t')$. Here, $\sigma(t)$ denotes the instantaneous direction of the active speed, which is given by a telegraphic or dichotomous noise that alternates between two values, $+1$ and $-1$, at a rate of $r_{\rm F}$. The particle begins from an initial position $x(0) = x_0$ with a specified direction $\sigma(0) = \sigma_0$. During its motion, the RTP's position is instantaneously reset to $x_{\rm R}$ at random times with a resetting rate of $r_{\rm R}$. When the particle's position is reset, the direction of the active speed may flip with a probability of $\eta$~\cite{Evans18}.

When the position $x(t)$ and the direction $\sigma(t)$ of the RTP's active speed at time $t$ are given, their subsequent values after an infinitesimal time interval $\Delta t$ are determined by one of the following four cases: (i) the position is reset, and the direction of active speed is reversed with probability $q_1 = r_{\rm R} \eta \Delta t$, (ii) the position is reset, but the direction of active speed remains unchanged with probability $q_2 = r_{\rm R} (1-\eta) \Delta t$, (iii) the motion of the particle  follows Eq.~\eqref{eq:RTP_equation} without direction flipping of active speed with probability $q_3 = (1- r_{\rm R} \Delta t)(1- r_{\rm F} \Delta t)$, and (iv) the motion of the particle  follows Eq.~\eqref{eq:RTP_equation} with direction flipping of active speed with probability $q_4 = (1- r_{\rm R} \Delta t)r_{\rm F} \Delta t$. In short, 
\begin{align}
     x(t+\Delta t) =  
        \left\{\begin{array}{llll}
		x_{\rm R} \hspace{0.2 cm}\&\hspace{.2 cm} \sigma(t)f\rightarrow -\sigma(t)f, & \text{with prob. $q_1$} \\
        x_{\rm R} \hspace{0.2 cm}\& \hspace{.2 cm} \sigma(t)f\rightarrow \sigma(t) f, & \text{with prob. $q_2$} \\
		x(t)+ \sigma(t) f\Delta t +\xi(t)\Delta t , & \text{with prob. $q_3$} \\
        x(t)-\sigma(t) f\Delta t+\xi(t)\Delta t , & \text{with prob. $q_4$ .}
	\end{array} \right. \label{eq:model_eq}
\end{align}
The diffusion and resetting process continues until the particle reaches the target position, which is set to the origin without loss of generality.
It is worth noting that previously introduced models can be derived as specific cases of our model by taking proper limits. For instance, when $f\rightarrow 0$, our model reduces to the original stochastic-resetting model~\cite{evans2011}. In another limit $D\rightarrow 0$, our model simplifies to one with only telegraphic noise~\cite{Evans18}.

\section{Moment generating functions of First Passage Time}
\label{sec:mgf}
FPT depends on RTP's initial configuration, including its starting position $x_0$ and initial direction $\sigma_0$. Therefore, we define the conditional FPT $t_{\rm F}^{\sigma_0}$, representing the first time the particle reaches the target (the origin) starting from $x_0$ with an initial direction $\sigma_0=\pm 1$. Following the Brownian functional method~\cite{curr-sc,Singh,Dubey}, one can express the moment generating function of FPT, conditioned on the given initial position $x_0$ and initial direction $\sigma_0$ as
\begin{align}
    Q_{\sigma_0}(p|x_0)=\int_0^{\infty} dt^{\sigma_0}_{\rm F} e^{-pt^{\sigma_0}_{\rm F}}P_{\sigma_0}(t^{\sigma_0}_{\rm F}|x_0)\equiv \la e^{-pt^{\sigma_0}_{\rm F}}\ra, 
    \label{def:gfpositive}    
\end{align}
where $P_{\sigma_0}(t^{\sigma_0}_{\rm F}|x_0)$ is the probability density function of $t^{\sigma_0}_{\rm F}$ for the process with $x(0) = x_0$ and $\sigma(0)=\sigma_0$. Then, the $m$th moment of $t_{\rm F}^{\sigma_0}$ can be evaluated as
\begin{equation} \label{eq:mth_moment}
    \langle (t_{\rm F}^{\sigma_0})^m \rangle =  (-1)^m \partial_p^m Q_{\sigma_0} (p|x_0) |_{p\rightarrow 0}. 
\end{equation}

Now we derive the differential equation for $Q_{\sigma_0}(p|x_0)$. Consider a trajectory of RTP whose FPT is denoted by $t^{\sigma_0}_{\rm F}$. We divide the trajectory into an initial infinitesimal segment with duration $\Delta t$ and the remaining part with duration $t^{\sigma_0}_{\rm F}-\Delta t$. During the initial infinitesimal segment, one of the four events can occur with respective probabilities $q_i$ as presented in Eq.~\eqref{eq:model_eq}. Then, the moment generating function $Q_{\sigma_0} (p|x_0)$ can be written as
\begin{align} \label{eq:infinitesimal_division}
    Q_{\sigma_0}(p|x_0) &= e^{-p\Delta t}\la e^{-p(t^{\sigma_0}_{\rm F}-\Delta t)}\ra \nn\\
    &=e^{-p\Delta t}\bigg\la  Q_{-\sigma_0}(p|x_{\rm R}) q_1 + Q_{\sigma_0}(p|x_{\rm R}) q_2 \bigg\ra\nn\\
    &+ e^{-p\Delta t}\bigg\la Q_{\sigma_0}(p|x_0') q_3 + Q_{-\sigma_0}(p|x_0'') q_4 \bigg\ra, 
\end{align}
where $x_0' = x_0 + \sigma_0 f \Delta t +\xi(0)\Delta t$ and $x_0'' = x_0 - \sigma_0 f\Delta t+\xi(0)\Delta t$.
By expanding Eq.~\eqref{eq:infinitesimal_division} and keeping the terms up to the order of $\Delta t$, we obtain the following backward differential equation of the moment generating functions: 
\begin{align} \label{DEQ:Q_diff_eq}
    \textbf{L}(\partial_{x_0})\textbf{Q}(p|x_0)=-r_{\rm R}\textbf{HQ}(p|x_{\rm R}),
\end{align}
where the moment generating functions are expressed as the vector $\textbf{Q}(p|x) = (Q_+(p|x), Q_-(p|x) )^{\textsf T}$ with $\textsf T$ denoting matrix transpose. The matrix $\textbf{L}(\hat{O})$ and $\textbf{H}$ are given by
\begin{align}
    \textbf{L}(\hat{O})=\begin{pmatrix} \mathcal{L}_{+}(\hat{O}) &r_{\rm F}\\r_{\rm F} & \mathcal{L}_{-}(\hat{O}) \end{pmatrix}~~\textrm{and}~~
    \textbf{H}=\begin{pmatrix}1-\eta & \eta \\ \eta & 1-\eta \end{pmatrix}, 
\end{align}
respectively, where $\mathcal{L}_{\pm}(\hat{O}) \equiv D \hat{O}^2 \pm f \hat{O}-g$ and $g\equiv p+r_{\rm R}+r_{\rm F}$.
To solve Eq.~\eqref{DEQ:Q_diff_eq}, we need to establish appropriate boundary conditions for $Q_{\sigma_0}(p|x_0)$. The first condition is that when $x_0 \rightarrow 0$, the RTP reaches the target instantly. The second condition requires that the moment generating functions converge to a finite value as $x\rightarrow \pm\infty$ due to instantaneous resetting. These boundary conditions can be expressed as:
\begin{align} \label{eq:bc}
    Q_{\pm}(p|x_0\rightarrow 0)=1,~~ Q_{\pm}(p|x_0\rightarrow \pm \infty)<\infty.
\end{align}
Note that $Q_{\sigma_0}(p|x_0)$ is symmetric in $x_0$ when $x_0$ is set to $x_{\rm R}$. In this case, it is sufficient to focus on the positive range of $x_0$.

We solve the differential equation~\eqref{DEQ:Q_diff_eq} analytically using the boundary conditions~\eqref{eq:bc}. In particular, when the initial position and the resetting position are the same, i.e., $x_0=x_{\rm R}$, the solutions can be expressed in the following compact form:
\begin{align} \label{sol_mgf}
    \textbf{Q}(p|x_{\rm R}) =\bigg[\mathbb{1}+r_{\rm R}(\mathbb{1}-\textbf{T}_{x_{\rm R}})\textbf{L}_0^{-1}\textbf{H}\bigg]^{-1}\textbf{T}_{x_{\rm R}}\textbf{I},
\end{align}
where $\mathbb{1}$ denotes the identity matrix, $\textbf{L}_0 = \textbf{L}(0)$, $\textbf{I} = (1,1)^{\textsf{T}}$, and  $\textbf{T}_{x} = \textbf{F} \textbf{E}_x \textbf{F}^{-1}$. Here, the matrices $\textbf{F}$ and $\textbf{E}_x$ are defined in Eq.~\eqref{eqA:FEC}. The derivation of Eq.~\eqref{sol_mgf} can be found  in Appendix~\ref{SecA:soln_for_MGF}.

\section{Results}
\label{sec:results}
\subsection{Unconditional mean first passage time}
\label{sec:unconditional_FTP}

To evaluate the  mean conditional FPT, we can
use Eq.~\eqref{eq:mth_moment} and the expressions of moment generating functions from Eq.~\eqref{sol_mgf} so that
\begin{align}
    \la t^{\pm}_{\rm F}\ra=-\partial_pQ_{\pm}(p|x_{\rm R})|_{p\rightarrow 0}.
    \label{fpt}
\end{align}
In real experiments, the initial direction of active speed is not fixed, but is determined probabilistically. Let $p_\pm$ represent the probabilities that the initial direction of active speed is $\sigma_0=\pm 1$; then the unconditional mean FPT can be calculated as
\begin{align}
    \la t_{\rm F} \ra = p_+ \la t^{+}_{\rm F} \ra +p_- \la t^{-}_{\rm F} \ra. 
    \label{def:uncon_fpt}
\end{align}
If the active particle is unbiased, it is reasonable to set $p_+ = p_- = 1/2$. This assumption is maintained throughout the rest of this paper. Hereafter, the unconditional mean FPT with $p_{\pm} = 1/2$ will be referred to simply as  mean FPT.

\begin{figure}
    \centering	\includegraphics[width=0.48\textwidth]{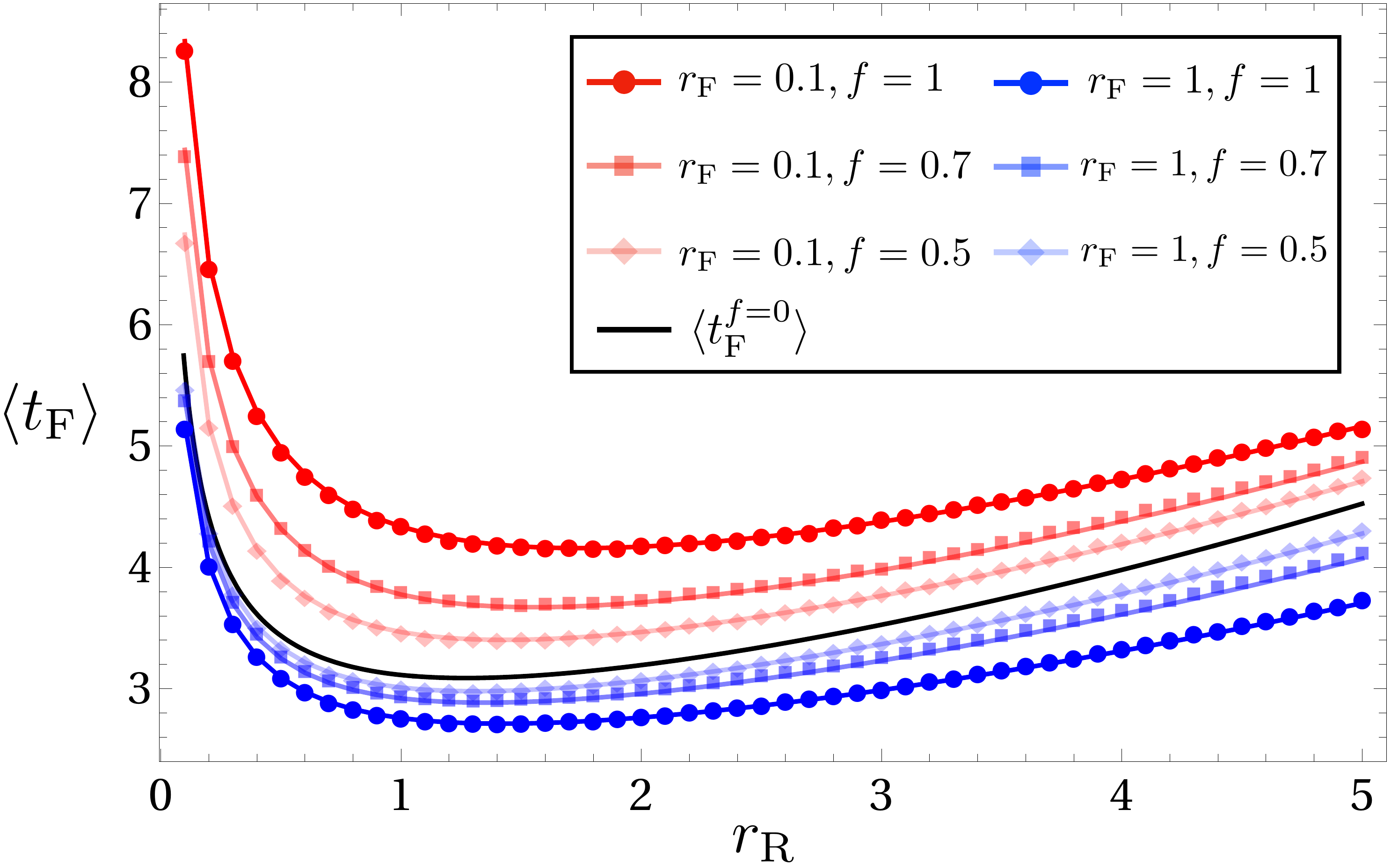}
    \caption{Plot of mean FPT $\la t_{\rm F} \ra$ as a function of resetting rate $r_{\rm R}$ for different combinations of flipping rate $r_{\rm F}$ and active speed $f$ at $\eta =0$ and $D=1/2$. Dots and solid curves represent simulation data and analytic results, respectively. }
    \label{fig:avgFPT_vs_rR}
\end{figure}

The full analytical expression for mean FPT is too lengthy to effectively handle and comprehend for physical insights. Instead, we illustrate the behavior of the mean FPT through various plots and discuss the impact of active motion on it. Figure~\ref{fig:avgFPT_vs_rR} shows the plot of $ \la t_{\rm F}\ra$ versus $r_{\rm R}$ for different combinations of flipping rate $r_{\rm F}$ and active speed $f$ at $D=1/2$. The various dots represent simulation data averaged over $10^5$ realizations of the explicit dynamics in Eq.~\eqref{eq:model_eq} and the solid curves indicate analytical results. For the simulation, $\Delta t$ in Eq.~\eqref{eq:model_eq} is set to be $10^{-5}$. The agreement between them validates our analysis. For this plot, reversing the direction of active speed during positional reset is not considered, i.e., $\eta$ is set to zero. The effect of a finite $\eta$ will be discussed in Section~\ref{sec:effect_noise_eta}. The value of $x_{\rm R}$ is consistently set to $1$ throughout this paper.

The black solid curve in Fig.~\ref{fig:avgFPT_vs_rR} shows mean FPT of the stochastic-resetting search using a  Brownian particle. This original process is identical to our model with $f=0$, resulting in the mean FPT denoted by $\la t_{\rm F}^{f=0}\ra$~\cite{evans2011}: 
\begin{equation} \label{eq:meanFPT_original}
    \la t_{\rm F}^{f=0}\ra = \frac{1}{r_{\rm R}}\left(e^{x_{\rm R}\sqrt{r_{\rm R}/D}}-1\right).       
\end{equation}
Compared to this original dynamics, when $r_{\rm F}=1$, mean FPT is shorter than $\la t_{\rm F}^{f=0}\ra$ and decreases as $f$ increases. This clearly demonstrates that active motion can reduce mean FPT compared to that of a Brownian particle. However, active motion is not always beneficial for enhancing search performance. When $r_{\rm F} = 0.1$, mean FPT becomes longer than $\la t_{\rm F}^{f=0}\ra$ and increases as $f$ rises. Regardless of magnitude of $f$ and $r_{\rm F}$, mean FPT exhibits a nonmonotonic behavior with respect to $x_{\rm R}$, implying that it is minimized at an optimal resetting rate $r_{\rm R}^*$. In particular, the optimal resetting rate for the $f=0$ case is denoted by $r_{{\rm R},f=0}^*$.

\begin{figure}
    \centering
    \includegraphics[width=0.48\textwidth]{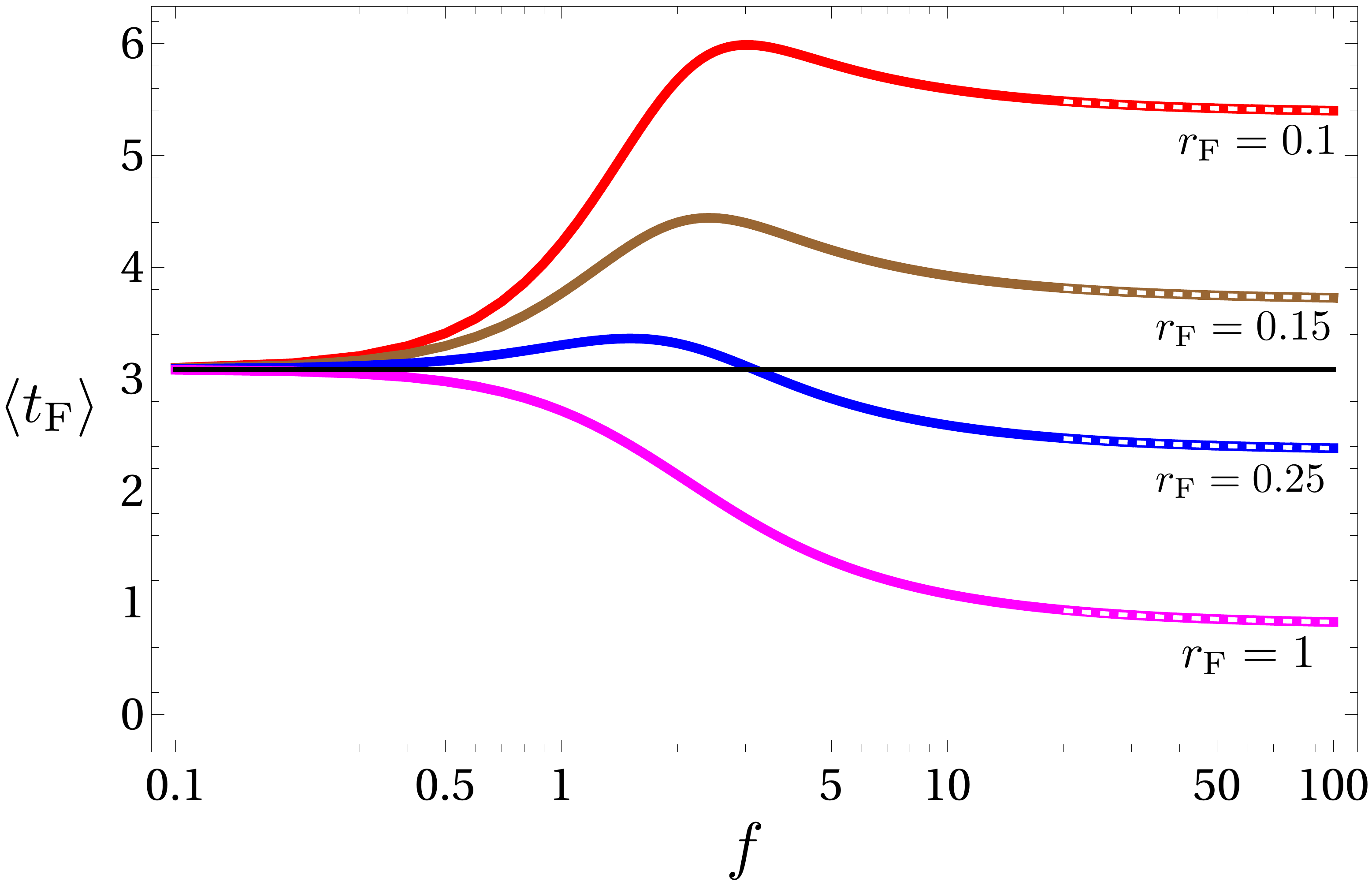}
    \caption{Plot of mean FPT as a function of  active speed $f$ for different values of flipping rate $r_{\rm F}$ at $\eta = 0$ and $D=1/2$. For all cases, resetting rate is set to $r_{{\rm R},f=0}^*$, which minimizes mean FPT of the original dynamics using a Brownian particle. Black solid line denotes mean FPT of the dynamics with $f=0$. Dashed curves represent the asymptotic behaviors of $\la t_{\rm F}\ra$ calculated from Eq.~\eqref{eq:asymptotic_FPT}. }
    \label{fig:avgFPT_vs_f_rRrME}
\end{figure}

To better understand the effect of active motion on mean FPT, we investigate dependence of $\la t_{\rm F}\ra$ on active speed and flipping rate at a fixed resetting rate. Figure~\ref{fig:avgFPT_vs_f_rRrME} shows the plot of $\la t_{\rm F}\ra$ versus $f$ for different values of $r_{\rm F}$ at $\eta = 0$ and $D=1/2$. For all curves in this plot, we set the resetting rate to $r_{\rm R}=r_{{\rm R},f=0}^*$ to demonstrate when $\la t_{\rm F}\ra$ is lower than the minimum mean FPT achieved by a Brownian particle. This minimum value, $\la t_{\rm F}^{f=0} \ra |_{r_{\rm R}=r_{{\rm R},f=0}^*}$, is indicated by the black line. 
As mentioned earlier, when active speed approaches zero, i.e., $f\rightarrow 0$, our model simplifies to the original model regardless of $r_{\rm F}$ value. This is evident in Fig.~\ref{fig:avgFPT_vs_f_rRrME}, where all curves converge to the black line for small $f$. In Appendix~\ref{secA:approximation}, we demonstrate that the leading term of mean FPT in the small $f$ expansion is independent of flipping rate and is expressed by 
\begin{align}
    \la t_{\rm F}\ra = \la t_{{\rm F},f=0}\ra+ {O}(f^2).
\end{align} 
In addition, we obtain the expression of mean FPT in the large $f$ limit as follows: 
\begin{align} \label{eq:asymptotic_FPT}
    \la t_{\rm F}\ra=\frac{\sqrt{1+\frac{2r_{\rm F}}{r_{\rm R}}}}{2(r_{\rm F}+\eta r_{\rm R})}+x_{\rm R}\bigg(1+\sqrt{1+\frac{2r_{\rm F}}{r_{\rm R}}}\bigg)\frac{1}{f} + {O}(f^{-2}).
\end{align}
Details of the derivation are provided in Appendix~\ref{secA:approximation}. This asymptotic behavior of $\la t_{\rm F}\ra$ in the large $f$ regime is represented by the dashed curves in Fig.~\ref{fig:avgFPT_vs_f_rRrME}. The dashed curves demonstrate that mean FPT saturates to a certain value at large $f$. This saturation value, given by $[2(r_{\rm F}+\eta r_{\rm R})]^{-1}\sqrt{1+2r_{\rm F}/r_{\rm R}}$, is a monotonically decreasing function of $r_{\rm F}$ for a fixed $r_{\rm R}$, indicating a higher $r_{\rm F}$ accelerates the search process. For moderate values of $f$, mean FPT decreases monotonically for high $r_{\rm F}$, while it exhibits nonmonotonic behavior for low $r_{\rm F}$. Nontheless, within this moderate range of $f$, mean FPT tends to decrease monotonically as $r_{\rm F}$ increases. Overall, increasing $r_{\rm F}$ highly seems to be a crucial factor to reduce $\la t_{\rm F}\ra$ below $\la t_{{\rm F},f=0}\ra$ across the entire range of $f$. 

This behavior of mean FPT can be understood in the following way. Active motion is characterized by two factors: run and tumble. Here, `run' or ballistic motion is governed by the parameter $f$, while `tumble' or stochasticity is controlled by the parameter $r_{\rm F}$. For a searching process where $x_{\rm R} = x_0 > 0$ and $\eta =0$, we first consider the case of low $r_{\rm F}$. At $f=0$, mean conditional FPTs with different initial directions are equal, i.e., $\la t_{\rm F}^+\ra = \la t_{\rm F}^-\ra$. However, if we increase $f$ from zero, the particle exhibits near-ballistic motion due to a finite $f$ with infrequent flipping. As a result, $\la t_{\rm F}^+\ra$ increases as the particle continues to move away from the target, while $\la t_{\rm F}^-\ra$ decreases as it approaches the target. 
In the large $f$ regime, it is straightforward to see that $\la t_{\rm F}^-\ra \rightarrow x_{\rm R}/f$. Moreover, $\la t_{\rm F}^+ \ra$ approaches $1/r_{\rm F} + x_{\rm R}/f$, where the first term, $r_{\rm F}^{-1}$, denotes the duration before the initial $+1$ direction of active speed flips into $-1$ direction, and the second term, $x_{\rm R}/f$, represents the time required to reach the target with $-1$ direction from $x_{\rm R}$. Thus, the mean unconditional FPT $\la t_{\rm F} \ra = (\la t_{\rm F}^+ \ra + \la t_{\rm F}^- \ra)/2 $ approaches $ 1/(2r_{\rm F}) + x_{\rm R}/f$, which is consistent with Eq.~\eqref{eq:asymptotic_FPT} for $r_{\rm F} \ll r_{\rm R}$. Therefore, for low value of $r_{\rm F}$, $\la t_{\rm F}\ra$ roughly increases from the value of $\la t_{\rm F}^{f=0}\ra$ to $1/(2r_{\rm F})$ as shown in Fig.~\ref{fig:avgFPT_vs_f_rRrME}. 

Now, we consider the case with high $r_{\rm F}$. In this case, active motion is more stochastic rather than ballistic. This stochastic tumbling of active speed effectively enhances the `diffusivity' of the particle, enabling it to explore the search area more rapidly. This effective diffusivity increases as $f$ increases. As a result, increasing $f$ expedites the search process and results in reducing FPT. This explains why we can achieve a lower mean FPT value compared to $\la t_{\rm F}^{f=0}\ra |_{r_{\rm R}=r_{{\rm R},f=0}^*}$ at high $r_{\rm F}$.

\subsection{Effects of thermal noise and finite $\eta$}
\label{sec:effect_noise_eta}

\begin{figure*}
    \centering
    \includegraphics[width=\textwidth]{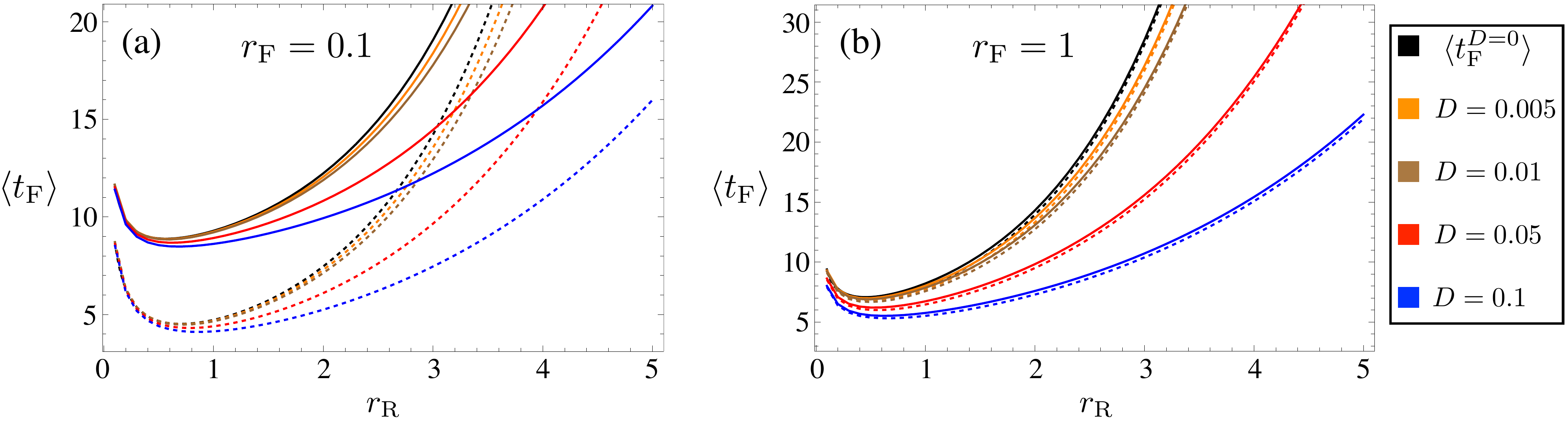}
    \caption{Effects of thermal noise and $\eta$ on mean FPT of RTP. $\la t_{\rm F}\ra$ is plotted as a function of $r_{\rm R}$ for different values of $\eta$ and $D$. (a) and (b) show plots with $r_{\rm F} = 0.1$ and $r_{\rm F} = 1$, respectively. For both plots, $f=1$ is used. The solid curves represent results with $\eta = 0$, where there is no flipping direction of active speed during reset. The dashed curves show results with $\eta = 1/2$, indicating that positional reset is accompanied by flipping direction of active speed with a probability $1/2$. Different colors represent different values of $D$, with black indicating dynamics without thermal noise (i.e., $D=0$). } 
    \label{fig:avgFPT_vs_rR_rF1_01}
\end{figure*}

We also explore the effects of thermal noise and finite $\eta$ on mean FPT. For a low flipping rate of $r_{\rm F} = 0.1$, the results are plotted in Fig.~\ref{fig:avgFPT_vs_rR_rF1_01}(a), which shows the mean FPT as a function of $r_{\rm R}$ for various $D$ and $\eta$. In the figure, solid curves represent $\la t_{\rm F}\ra$ with $\eta = 0$ for different $D$ values. Among these, the black solid curve indicates mean FPT for RTP without thermal noise, i.e., the $D = 0$ case. The analytic expression of mean FPT for RTP with $D = 0$, denoted by $\la t_{\rm F}^{D=0}\ra$, is given by~\cite{Evans18}
\begin{align} \label{eq:t_F_zeroD}
    \la t_{\rm F}^{D=0}\ra
    =\left\{\begin{array}{ll}
    -\frac{1}{2r_{\rm F}R}+\frac{e^{\chi\sqrt{R(1+R)}}-R^2+R\sqrt{R(1+R)}}{2r_{\rm F}R[1+R-\sqrt{R(1+R)}]}, & \eta=0 \\
    -\frac{1}{2r_{\rm F}R}+\frac{e^{\chi\sqrt{R(1+R)}}}{2r_{\rm F}R[1+R-\sqrt{R(1+R)}]}, & \eta=\frac{1}{2} \\	
\end{array} \right.    
\end{align}
where $R \equiv r_{\rm R}/(2r_{\rm F})$ and $\chi \equiv 2r_{\rm F}x_{\rm R}/f$. Compared to $\la t_{\rm F}^{D=0}\ra$, mean FPT with a finite $D$ is always lower and decreases as $D$ increases as the figure shows. Dashed curves in Fig.~\ref{fig:avgFPT_vs_rR_rF1_01}(a) represent mean FPT with $\eta=1/2$ for different $D$ values. As also seen from Eq.~\eqref{eq:t_F_zeroD}, $\la t_{\rm F}^{D=0}\ra$ with $\eta=1/2$ is lower than that with $\eta=0$. Moreover, increasing $D$ from $D=0$ further reduces mean FPT. Similar to the case of a low flipping rate, for a high flipping rate $r_{\rm F}=1$, $\la t_{\rm F}\ra$ decreases as $D$ or $\eta$ increases, as shown in Fig.~\ref{fig:avgFPT_vs_rR_rF1_01}(b). 

These results indicate that thermal noise and flipping of  active speed during reset expedite the searching process. We can understand these behaviors as follows: First, increasing $D$ raises diffusivity, thus helping the particle explore the searching area more rapidly. Consequently, this results in a reduced mean FPT. This principle is also valid for the original model using a Brownian particle, where mean FPT, expressed as~\eqref{eq:meanFPT_original}, decreases as $D$ increases. 
Second, an increase in $\eta$ is effectively similar to raising $r_{\rm F}$, which also aids in reducing mean FPT, as discussed in Section~\ref{sec:unconditional_FTP}. This effect is evident from the asymptotic expression of mean FPT in the large $f$ regime in Eq.~\eqref{eq:asymptotic_FPT}; the saturated mean FPT, $[2(r_{\rm F}+\eta r_{\rm R})]^{-1}\sqrt{1+2r_{\rm F}/r_{\rm R}}$, is a monotonically decreasing function of $\eta$. Particularly, the impact of increasing $\eta$ is more pronounced for low $r_{\rm F}$, as seen in the expression and the plots shown in Figs.~\ref{fig:avgFPT_vs_rR_rF1_01}(a) and \ref{fig:avgFPT_vs_rR_rF1_01}(b).

\subsection{Optimal resetting rate and beneficiality of active motion}
\label{sec:optimal}
In this section, we provide a comprehensive analysis of when an RTP outperforms a Brownian particle as a search agent. Suppose there exists a Brownian particle in a thermal environment with a diffusion constant $D$ and an RTP with active speed $f$ and flipping rate $r_{\rm F}$ in the same thermal bath. If mean FPT of the RTP is lower than that of the Brownian particle, one can say that using an RTP is more beneficial for searching a target than a Brownian particle. For a given particle, however, mean FPT varies depending on $r_{\rm R}$. Specifically, mean FPT is a nonmonotonic (convex) function of the resetting rate, and thus, it is minimized at an optimal value of $r_{\rm R}=r_{\rm R}^*$. The optimal resetting rate for RTP $r_{\rm R}^*$ is a function of $f$, $r_{\rm F}$, and $D$, whereas the optimal resetting rate for a Brownian particle $r_{{\rm R},f=0}^*$ depends solely on $D$. Therefore, it is reasonable to compare the minimum mean FPTs evaluated at $r_{\rm R}^*$ for RTP and $r_{{\rm R},f=0}^*$ for Brownian particle, to quantitatively assess the outperformance of RTP. In this context, as a measure for discerning the degree of benefit of using an active particle for searching a target, we introduce the concept of \emph{beneficiality} of active motion $\mathcal{B} = \mathcal{B}(D, f, r_{\rm F})$ as a logarithmic ratio between the minimum mean FPTs of an RTP $\la t_{\rm F}\ra |_{r_{\rm R}^{ } = r_{\rm R}^*}$ and a Brownian particle $\la t_{\rm F}^{f=0}\ra |_{r_{\rm R}^{} = r_{{\rm R},f=0}^*}$:
\begin{equation}
    \mathcal{B}=\log\bigg[\frac{\la t_{\rm F}\ra|_{r_{\rm R}=r_{\rm R}^*}}{\la t_{\rm F}^{f=0}\ra |_{r_{\rm R} = r_{{\rm R},f=0}^*}}\bigg].    
\end{equation}
Therefore, when $\mathcal{B} > 0$, active motion provides no benefit for searching compared to a Brownian particle. Conversely, when $\mathcal{B} < 0$, an active particle outperforms a Brownian particle.

\begin{figure*}
	\centering
		\includegraphics[width=\textwidth]{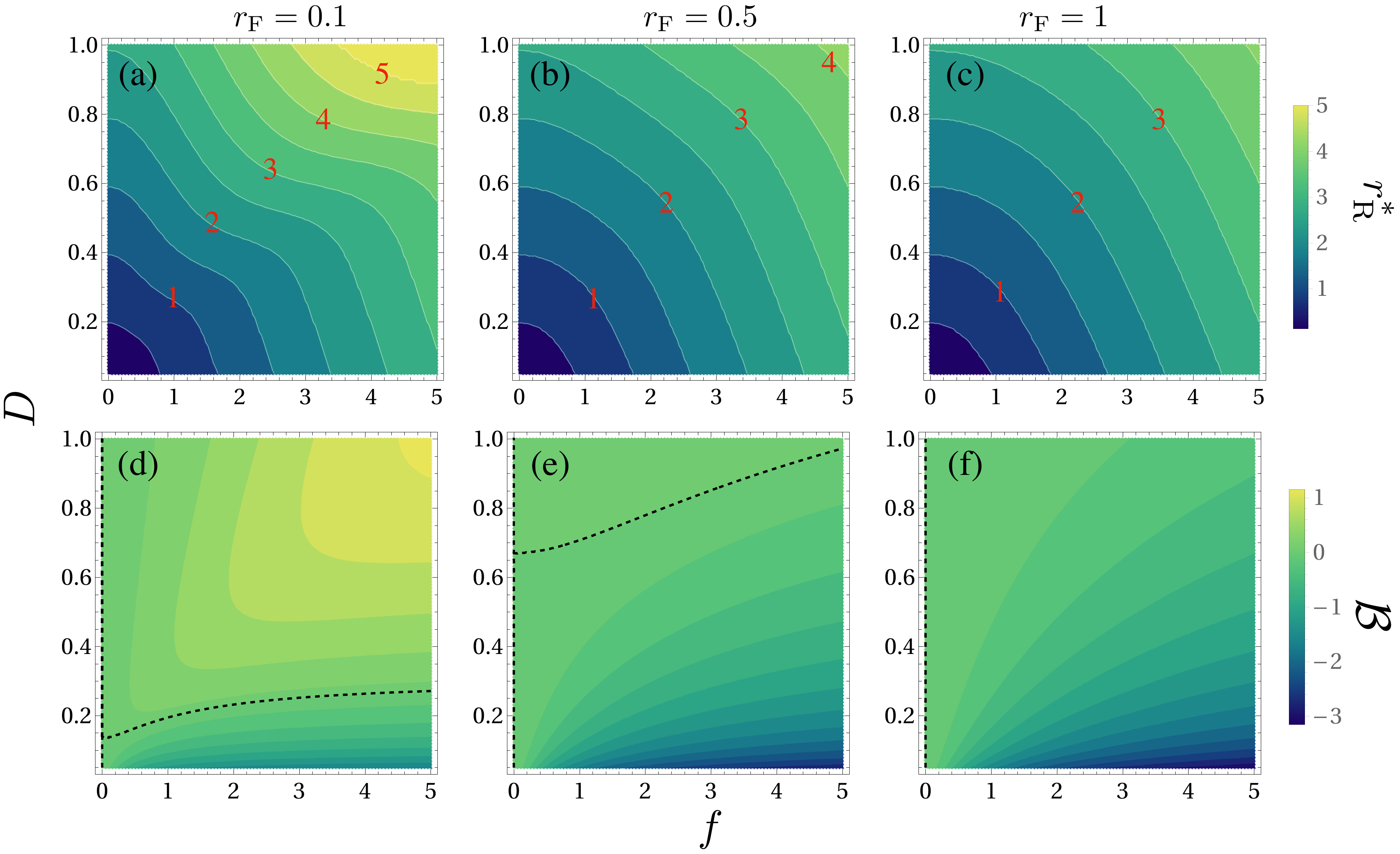}
		\caption{Panels (a), (b), and (c) depict the contour plots of the optimal resetting rate $r_{\rm R}^*$ as a function of diffusion coefficient $D$ and active speed $f$ for three different values of flipping rate $r_{\rm F}=$ $0.1$ , $0.5$, and $1$, respectively. Panel (d), (e), and (f) are the contour plots of beneficiality $\mathcal{B}$ as a function of $D$ and $f$ for three different values of flipping rate $r_{\rm F}=$ $0.1$ , $0.5$, and $1$, respectively. Black dashed lines denote isolines where $\mathcal{B}=0$.}
\label{fig:rRstar_logratiofpt_vs_Df}
\end{figure*}

To evaluate $\mathcal{B}$, we begin by estimating the optimal resetting rate $r_{\rm R}^*$ numerically from data of $\la t_{\rm F}\ra$ as a function of $r_{\rm R}$ for a wide range of combinations of $D$, $f$, and $r_{\rm F}$.  Figures~\ref{fig:rRstar_logratiofpt_vs_Df}(a), \ref{fig:rRstar_logratiofpt_vs_Df}(b), and \ref{fig:rRstar_logratiofpt_vs_Df}(c) present two-dimensional contour plots showing how $r_{\rm R}^*$ varies with $D$ and $f$ for three different values of $r_{\rm F}$: $0.1$, $0.5$, and $1$. These plots also include isolines where $r_{\rm R}^*(D, f, r_{\rm F})$ remains constant. Overall, the tendency is for $r_{\rm R}^*$ to increase as either $D$ or $f$ increases.

Using this estimated $r_{\rm R}^*$, we evaluate $\mathcal{B}$. Figures~\ref{fig:rRstar_logratiofpt_vs_Df}(d)-(f) show contour plots of $\mathcal{B}$ versus $D$ and $f$ for different values of $r_{\rm F}$. The black dashed lines represent isolines where $\mathcal{B}=0$. These isolines divide the $D$-$f$ space into two regions: (i) the $\mathcal{B}>0$ region, where active motion provides no benefit, and (ii) the $\mathcal{B}<0$ region, where active particle outperforms a Brownian particle.

From these observation, we can deduce the overall tendencies of $\mathcal{B}$ as follows. First, the region with $\mathcal{B}<0$ expands as $r_{\rm F}$ increases, suggesting that higher $r_{\rm F}$ is advantageous for enhancing search performance. Second, active motion loses its benefit as $D$ increases. This results from the fact that the decrease in $\la t_{\rm F}^{f=0}\ra |_{r_{\rm R}^{} = r_{{\rm R},f=0}^*}$ caused by increasing $D$ outweighs the decrease in $\la t_{\rm F}\ra |_{r_{\rm R}^{ } = r_{\rm R}^*}$. This highlights the importance of environment; active motion is most beneficial in low-temperature environments. Third, increasing $f$ usually aids in reducing mean FPT, except in cases with high $D$ and low $r_{\rm F}$. In summary, high $r_{\rm F}$ and large $f$ in an environment with low $D$ are best conditions for achieving maximum benefit from an RTP.

\section{Conclusion}
\label{sec:conclusion}

We have investigated the target-search problem using an active particle governed by run-and-tumble dynamics in the presence of a thermal noise under the instantaneous stochastic resetting mechanism. Our model is a generalization of two previously studied stochastic resetting models: (i) simple diffusive search in a thermal bath~\cite{evans2011} and (ii) simple run-and-tumble search in a non-thermal environment~\cite{Evans18}. Most of the earlier results can be retrieved from ours when appropriate limits are taken. From our model, we find that active motion significantly modifies the search dynamics either in reducing or prolonging the mean FPT. To understand the overall combined influence of active motion and thermal noise on mean FPT, we have introduced the function `beneficiality' as a measure for discerning the degree of benefit of using an active particle for searching a target. Extensive calculations verify that large $f$, high $r_{\rm F}$, and low $D$ are the best choices for an active particle to outperform a Brownian particle as a search agent. We also demonstrate that the thermal environment plays a significant role in reducing mean FPT. In addition to active motion and thermal noise, the search process by an RTP can also be made more effective by reversing the direction of the active speed during position resetting.

It is worth noting that our model enables the conducting of target-searching experiments using RTP in a thermal environment. Previous studies solely considered RTP without thermal noise~\cite{Evans18}, making it experimentally difficult to prepare such conditions because thermal noise is indispensable in mesoscopic systems. Moreover, in the real world, especially in biological systems, search processes can take place in more complex environments. Therefore, it is crucial to investigate the combined effect of active motion and environmental influences on searching dynamics. In this regard, our approach and the proposed measure of `beneficiality' will be a good starting point and basis for future research in this direction.

\section{Acknowledgement}
P.S.P. and J.-M.P. equally contributed to this work. The authors acknowledge Korea Institute for Advanced Study for providing computing resources [KIAS Center for Advanced Computation Linux Cluster System]. This research was supported by NRF Grant No.~2017R1D1A1B06035497 (H.P.), and individual KIAS Grants No.~PG064902 (J.S.L.), No.~PG085601 (P.S.P), and No.~QP013601 (H.P.) at Korea Institute for Advanced Study. AP gratefully acknowledges research support from the DST-SERB Start-up Research Grant Number SRG/2022/000080 and the DAE, Govt. of India.


\begin{appendix}
\renewcommand{\theequation}{A\arabic{equation}}
\renewcommand{\thefigure}{A\arabic{figure}}
\setcounter{equation}{0}
\setcounter{figure}{0}

\section{Solution of Eq.~\eqref{DEQ:Q_diff_eq}}
\label{SecA:soln_for_MGF}

Solution of Eq.~\eqref{DEQ:Q_diff_eq} can be attained by dividing $\textbf{Q}(p|x_0)$ into the homogeneous and inhomogeneous parts as
\begin{align}
    \textbf{Q}(p|x_0)=\textbf{Q}^{\rm h}(p|x_0)+\textbf{Q}^{\rm inh}(p|x_{\rm R}).
\end{align}
Then, the inhomogeneous part satisfies the following equation:
\begin{align} \label{eqA:Q_inh}
    \textbf{L}_0 \textbf{Q}^{\rm inh}(p|x_{\rm R})&=-\textbf{HR}~ \implies ~\textbf{Q}^{\rm inh}(p|x_{\rm R})=-\textbf{L}_0^{-1}\textbf{HR},
\end{align}
where $\textbf{L}_0 \equiv\textbf{L}(0)$ and $\textbf{R}\equiv r_{\rm R}\textbf{Q}(p|x_{\rm R})$. The trial solution for the homogeneous part $\textbf{Q}^{\rm h}(p|x_0)=C\boldsymbol{\Phi}(p)e^{mx_0}$, where $C$ is a constant independent of $p$ and $x_0$, satisfies
\begin{align} \label{eqA:trial_eq}
    \textbf{L}(\partial_{x_0})C\boldsymbol{\Phi}(p)e^{mx_0}=\textbf{L}(m)C\boldsymbol{\Phi}(p)e^{mx_0}=0.
\end{align}
Nonzero solutions of Eq.~\eqref{eqA:trial_eq} requires $\text{det}[\textbf{L}(m)]=0$ which is expressed as
\begin{align} \label{eqA:m_eq}
    (Dm^2-g)^2-f^2m^2-r_{\rm F}^2&=0.
\end{align}
The four solutions of Eq.~\eqref{eqA:m_eq} are
\begin{align}
    m_{1,\pm}&=\pm \frac{\sqrt{2\ell+2\sqrt{\ell^2-4D^2(g^2-r_{\rm F}^2)}}}{2D},\nn\\
    m_{2,\pm}&=\pm \frac{\sqrt{2\ell-2\sqrt{\ell^2-4D^2(g^2-r_{\rm F}^2)}}}{2D},
\end{align}
where $\ell \equiv 2gD+f^2$. Note that all $m_{i,\pm}$ ($i=1,2$) are real numbers. First, $m_{1,\pm}$ are real because
\begin{align}
    \ell^2-4D^2(g^2-r_{\rm F}^2) = f^4+4gDf^2+4D^2r_{\rm F}^2 > 0. \nn
\end{align}
$m_{2,\pm}$ are also real numbers because
\begin{align}
    \ell^2 -\sqrt{\ell^2-4D^2(g^2-r_{\rm F}^2)}^2 = 4D^2(g^2-r_{\rm F}^2) > 0,
\end{align}
where $g=p+r_{\rm R}+r_{\rm F}$ with $p,r_{\rm R},r_{\rm F} > 0$. The corresponding eigenvectors are given by 
\begin{align}
    \textbf{L}(m)\boldsymbol{\Phi}(p)=0 ~
    \implies ~ \boldsymbol{\Phi}_{i,\pm}(p)=\begin{pmatrix}
        F_{i,\pm}-fm_{i,\pm}\\F_{i,\pm}+fm_{i,\pm}
    \end{pmatrix},
\end{align}
where $F_{i,\pm} \equiv Dm_{i,\pm}^2-g-r_{\rm F}$. Hence, the general homogeneous solution is expressed as
\begin{align}
    &\textbf{Q}^{\rm h} (p|x_0) \nn\\
    &=\sum_{i=1}^2 [C_{i,+}\boldsymbol{\Phi}_{i,+}(p)e^{m_{i,+}x_0}+C_{i,-}\boldsymbol{\Phi}_{i,-}(p)e^{m_{i,-}x_0}].
\end{align}
The boundary condition $Q_{\pm}(p|x_0\rightarrow \infty)<\infty$ ensures $C_{i,+}=0$ for $i=1,2$ as $m_{i,+}>0$. As a result, the expression of the homogeneous part of the moment generating function can be rewritten as
\begin{align}
\textbf{Q}^{\rm h} (p|x_0) = \textbf{FE}_{x_0}\textbf{C},
\end{align}
where $\textbf{F}$, $\textbf{E}_{x}$, and $\textbf{C}$ are defined as follows:
\begin{align} \label{eqA:FEC}
    &\textbf{F}\equiv \begin{pmatrix}
        F_{1,-}-fm_{1,-} & F_{2,-}-fm_{2,-}\\F_{1,-}+fm_{1,-}&F_{2,-}+fm_{2,-}
    \end{pmatrix},\nn\\
    &\textbf{E}_{x}\equiv \begin{pmatrix}
        e^{m_{1,-} x}&0\\0&e^{m_{2,-}x}
    \end{pmatrix}, ~~\textbf{C}\equiv \begin{pmatrix}
        C_{1,-}\\C_{2,-}
    \end{pmatrix}.
\end{align}
Therefore, the solution of the moment generating function is
\begin{align} \label{eqA:Q_intermediate1}
    \textbf{Q}(p|x_0)=\textbf{FE}_{x_0}\textbf{C}+\textbf{Q}^{\rm inh}(p|x_{\rm R}).
\end{align}
Using the boundary condition $Q_{\pm}(p|x_0\rightarrow 0)=1$, we have
\begin{align}
    \textbf{C}=\textbf{F}^{-1}(\textbf{I}-\textbf{Q}^{\rm inh}),
\end{align}
where $\textbf{I}=\begin{pmatrix} 1, 1\end{pmatrix}^{\textsf T}$. Hence, Eq.~\eqref{eqA:Q_intermediate1} is rewritten as
\begin{align} \label{eqA:Q_intermediate2}
    \textbf{Q}(p|x_0)=\textbf{T}_{x_0}\textbf{I}+(\mathbb{1}-\textbf{T}_{x_0})\textbf{Q}^{\rm inh},
\end{align}
where $\textbf{T}_{x}=\textbf{FE}_{x}\textbf{F}^{-1}$. Plugging $\textbf{Q}(p|x_{\rm R}) = \textbf{R}/r_{\rm R}$ and $\textbf{Q}^{\rm inh}(p|x_{\rm R})=-\textbf{L}_0^{-1}\textbf{HR}$ in Eq.~\eqref{eqA:Q_inh} into Eq.~\eqref{eqA:Q_intermediate2}, we have 
\begin{align}
    \textbf{R} =\bigg[\frac{1}{r_{\rm R}}\mathbb{1}+(\mathbb{1}-\textbf{T}_{x_{\rm R}})\textbf{L}_0^{-1}\textbf{H}\bigg]^{-1}\textbf{T}_{x_{\rm R}}\textbf{I}.
\end{align}
Then, the inhomogeneous part of the moment generating function is
\begin{align}
    \textbf{Q}^{\rm inh}(p|x_{\rm R})=-\textbf{L}_0^{-1}\textbf{H}\bigg[\frac{1}{r_{\rm R}}\mathbb{1}+(\mathbb{1}-\textbf{T}_{x_{\rm R}})\textbf{L}_0^{-1}\textbf{H}\bigg]^{-1}\textbf{T}_{x_{\rm R}}\textbf{I}.
\end{align}
By substituting this $\textbf{Q}^{\rm inh}(p|x_{\rm R})$ expression into Eq.~\eqref{eqA:Q_intermediate2}, we finally arrive at
\begin{align}
    \textbf{Q}(p|x_0)=\textbf{T}_{x_0}\textbf{I}-\textbf{K}_{x_0} \bigg[\frac{1}{r_{\rm R}}\mathbb{1}+\textbf{K}_{x_{\rm R}}\bigg]^{-1}\textbf{T}_{x_{\rm R}}\textbf{I},
\end{align}
where $\textbf{K}_x \equiv (\mathbb{1}-\textbf{T}_{x})\textbf{L}_0^{-1}\textbf{H}$.
For simplicity, if we choose $x_0=x_{\rm R}$, the moment generating function simplifies as
\begin{align} \label{eqA:Q_soln}
    \textbf{Q}(p|x_{\rm R})&=r_{\rm R}\textbf{R}\nn\\
    &=\bigg[\mathbb{1}+r_{\rm R}(\mathbb{1}-\textbf{T}_{x_{\rm R}})\textbf{L}_0^{-1}\textbf{H}\bigg]^{-1}\textbf{T}_{x_{\rm R}}\textbf{I}.
\end{align}


\renewcommand{\theequation}{B\arabic{equation}}
\setcounter{equation}{0}
\renewcommand{\thefigure}{B\arabic{figure}}
\setcounter{figure}{0}

\section{Derivation of approximate expressions of mean FPT }
\label{secA:approximation}

Although the complete expression of mean FPT can be achieved from Eq.~\eqref{fpt}, it is too lengthy to handle and interpret for physical insights. Therefore, in this section, we present mean FPT in two limiting cases: small and large $f$. First, in the limit of small $f$, $m_{i,-}$ simplifies as
\begin{align}
    m_{1,-}&=-\sqrt{\frac{g+r_{\rm F}}{D}}+O(f^2),\nn\\
    m_{2,-}&=-\sqrt{\frac{g-r_{\rm F}}{D}}+O(f^2).
\end{align}
Furthermore, $\textbf{F}$ and $\textbf{E}_{x_{\rm R}}$ in the small $f$ limit are given by
\begin{align}
    &\textbf{F}=\begin{pmatrix}
        f\sqrt{\frac{g+r_{\rm F}}{D}}&-2r_{\rm F}+f\sqrt{\frac{g-r_{\rm F}}{D}}\\
        -f\sqrt{\frac{g+r_{\rm F}}{D}}&-2r_{\rm F}-f\sqrt{\frac{g-r_{\rm F}}{D}}
    \end{pmatrix}+O(f^2),\nn\\
    &\textbf{E}_{x_{\rm R}}=\begin{pmatrix}
        e^{-\sqrt{\frac{g+r_{\rm F}}{D}}x_{\rm R}}&0\\
        0& e^{-\sqrt{\frac{g-r_{\rm F}}{D}}x_{\rm R}}
    \end{pmatrix}+O(f^2).
\end{align}
Then, from Eqs.~\eqref{fpt} and \eqref{eqA:Q_soln}, the conditional mean FPT for small $f$ becomes
\begin{align} \label{eqA:meanFPT_smallf}
    \la t_{\rm F}^{\pm}\ra=\frac{1}{r_{\rm R}} \left(e^{x_{\rm R}\sqrt{\frac{r_{\rm R}}{D}}}-1 \right)\pm \mathcal{A}f+O(f^2),
\end{align}
where $\mathcal{A}$ is defined as 
\begin{align} \label{eqA:meanFPT_smallf}
    \mathcal{A} \equiv \frac{\left(e^{x_{\rm R}\sqrt{\frac{2r_{\rm F}+r_{\rm R}}{D}}}-e^{x_{\rm R}\sqrt{\frac{r_{\rm R}}{D}}} \right) (2r_{\rm F}+r_{\rm R}) }{2r_{\rm F}\sqrt{Dr_{\rm R}}\left(2r_{\rm F}e^{x_{\rm R}\sqrt{\frac{2r_{\rm F}+r_{\rm R}}{D}}}+r_{\rm R}\right)}.
\end{align}
Hence, in the limit of small $f$, the unconditional mean FPT approaches that of a freely diffusing Brownian particle as expected:
\begin{align}
    \la t_{\rm F}\ra =\frac{1}{2}(\la t_{\rm F}^+\ra+ \la t_{\rm F}^-\ra) =\frac{1}{r_{\rm R}}\left(e^{x_{\rm R}\sqrt{\frac{r_{\rm R}}{D}}}-1\right)+O(f^2).
\end{align}
Next, in the limit of large $f$ (or equivalently small $f^{-1}$), $m_{i,-}$ reduces to 
\begin{align}
    m_{1,-}&=-\frac{f}{D}-\frac{g}{f}+O(f^{-2}), \nn\\
    m_{2,-}&=-\frac{\sqrt{g^2-r_{\rm F}^2}}{f}+O(f^{-2}).
\end{align}
In addition, $\textbf{F}$ and $\textbf{E}_{x_{\rm R}}$ in the large $f$ limit are expressed as
\begin{align}
    &\textbf{F}=\begin{pmatrix}
         2g-r_{\rm F}+\frac{2}{D}f^2 & -g-r_{\rm F}+\sqrt{g^2-r_{\rm F}^2}\\
         -r_{\rm F} & -g-r_{\rm F}-\sqrt{g^2-r_{\rm F}^2}
    \end{pmatrix}+O(f^{-2}), \nn\\
    &\textbf{E}_{x_{\rm R}}=\begin{pmatrix}
         0 & 0\\
         0 & 1-\frac{x_{\rm R}\sqrt{g^2-r_{\rm F}^2}}{f}
    \end{pmatrix} +O(f^{-2}).
\end{align}
Thus, the conditional mean FPT in this limit is given by
\begin{align}
    \la t_{\rm F}^+\ra&=\frac{\sqrt{1+\frac{2r_{\rm F}}{r_{\rm R}}}}{r_{\rm F}+\eta r_{\rm R}} +x_{\rm R}\bigg(1+\sqrt{1+\frac{2r_{\rm F}}{r_{\rm R}}}\bigg)\frac{1}{f} +O(f^{-2}),\nn\\
    \la t_{\rm F}^-\ra&=x_{\rm R}\bigg(1+\sqrt{1+\frac{2r_{\rm F}}{r_{\rm R}}}\bigg)\frac{1}{f}+O(f^{-2}).
\end{align}
Finally, we arrive at the expression of the unconditional mean FPT for large $f$ as
\begin{align}
    \la t_{\rm F}\ra&=\frac{1}{2}(\la t_{\rm F}^+\ra+ \la t_{\rm F}^-\ra)\nn\\
     &=\frac{\sqrt{1+\frac{2r_{\rm F}}{r_{\rm R}}}}{2(r_{\rm F}+\eta r_{\rm R})}  +x_{\rm R}\bigg(1+\sqrt{1+\frac{2r_{\rm F}}{r_{\rm R}}}\bigg)\frac{1}{f}+O(f^{-2}).
\end{align}     


\end{appendix}

\end{document}